\documentclass[12pt]{article}
\usepackage[colorlinks=true]{hyperref}
\usepackage{amsmath}

\usepackage{amsfonts,amsxtra,graphicx}

\def\be{\begin{eqnarray}}
\def\ee{\end{eqnarray}}

\begin{document}

\centerline{\bf \Large Real or Imaginary?(On pair creation in de Sitter space)}

\vspace{5mm}

\centerline{{\bf Emil T. Akhmedov}\footnote{E--mail:akhmedov@itep.ru}} 

\vspace{5mm}

\centerline{ITEP, B. Cheremushkinskaya, 25, Moscow, Russia 117218}

\vspace{5mm}


\begin{abstract}
Using properly defined Feynman propagator we obtain
non--zero imaginary contribution to the scalar field effective action
in even dimensional de Sitter space. Such a propagator follows from the path
integral in de Sitter space and obeys composition principle proposed
in arXiv:0709.2899. The obtained expression for the effective action
shows particle production with the Gibbons--Hawking rate.
\end{abstract}


\section{Introduction}

There is a controversy in the literature (see e.g. \cite{Candelas:1975du}--\cite{Mottola:1984ar}) on the issue of the presence of the Schwinger type imaginary contributions in the one--loop scalar effective actions in de Sitter (dS) space and on the physical implications (see e.g. \cite{Mottola:1984ar}--\cite{Volovik:2009eb})
of the corresponding pair production \cite{Gibbons:1977mu}.

To calculate the imaginary contribution to the effective action one has to use
the in--out Feynman propagator. Usually one
uses the so called Bunch--Davies \cite{Bunch:1978yq} Feynman propagator, which
is constructed for the scalar, $\phi$, field theory as follows $G_{FBD}(x,y) =
\langle BD| T\{\phi(x)\, \phi(y)\}|BD\rangle$, where $|BD\rangle$ is the Bunch--Davies
vacuum. In this case one does not find any imaginary Schwinger type contribution
to the one loop effective action.

However, such a propagator does not obey the composition principle,
which should be obeyed by the in--out Feynman propagator\cite{Polyakov:2007mm}.
The composition principle is the fundamental property
of the in--out Feynman propagator $G_F(x,y)$ in QFT stating that $\int dy \, G_F(x,y)\,
G_F(y,z)|_{L(x,z)\to\infty} \propto G_F(x,z)$, where $L(x,y)$ is the distance between the two points
$x$ and $y$. This property follows from the following behavior of the propagator $G_F(x,y)\propto e^{-i m\,L(x,y)}$ at large distances $L(x,y)$ and should be obeyed to at least justify transition
of the QFT in question to its classical limit.  Bunch--Davies propagator does not obey
this principle because it behaves, at large distances $L(x,y)$, as $G_{BD}(x,y)\propto c \, e^{-i m\,L(x,y)}
+ c^* \, e^{i m\,L(x,y)}$, for some complex constant $c$.

In this note we show that the path integral in dS space leads to a propagator which obeys the composition principle. This propagator was proposed in \cite{Polyakov:2007mm}, where it is shown
that for the scalar field it can be constructed as $G_{F} = \langle out| T\{\phi(x)\, \phi(y)\}|BD\rangle/\langle out| BD\rangle$, where $|out\rangle$ is the so called out state
for QFT in dS space --- the state which is defined wrt harmonics which are diagonalizing the free
Hamiltonian in the future infinity.

We show that analytical continuation of the path integral
itself from the sphere to dS or to Euclidian anti--de Sitter (EAdS) space is ill defined.
At the same time, properly defined path integrals on dS and EAdS lead to the correct Feynman
propagators, which
obey composition principle and are related to each other via analytical continuation\footnote{The justification for the analytical continuation from dS to EAdS follows from the fact that under the
change \cite{Polyakov:2007mm} of the curvature
$R\to i\,R$ and the interval $ds \to i\, ds$
the metrics, geodesic equations and Klein--Gordon operators in dS and EAdS
spaces are mapped to each other.}.
The dS propagator defined in such a way leads to the non--zero finite imaginary contribution
to the one--loop effective action. This contribution shows the vacuum decay with the Gibbons--Hawking rate.

Our modest new contribution here is that we explicitly show that one can obtain the Polyakov's
propagator from the path integral in dS space; we show that this propagator and the corresponding
path integral are related by the analytical continuation to the correct ones in EAdS space;
we show that one can not obtain the path integral and the propagator in EAdS space via analytical
continuation from those on the sphere.

\section{Effective actions}

In this section we calculate the effective actions for different choices of scalar Feynman propagators in dS spaces. For the beginning we sketch the calculation for the $\alpha$--vacua. Our notations and details for the Green functions are given in the Appendix.

In this paper we take space--time dimension of dS space $D$
to be even and set the Hubble constant to be $H=1$.
To find the scalar effective Lagrangian $L_{eff}$ we use the relation \cite{Candelas:1975du}
$\frac{{\rm Im}\, L_{eff}}{\sqrt{|g|}} \propto i\, {\rm Re} \int_{+\infty}^{m^2} dm^2 \,G_F(z=1)$.
Here $G_F(z=1)$ is the value of the Feynman propagator at the coincident points.
The general scalar Feynman propagator corresponding to $\alpha$--vacua is

\begin{eqnarray}
\label{BD}
G_F(z) = \frac{i\,\left|\Gamma(h_+)\right|^2}{\Gamma\left(\frac{D}{2}\right)\, \Gamma\left(\frac{D}{2} - 1\right)} \,
\left[\cosh(\alpha) \,F\left(h_+,h_-;\frac{D}{2};\frac{1+z}{2}-i\,\epsilon\right) \right. \nonumber \\ +
\left. \sinh(\alpha)\, F\left(h_+,h_-;\frac{D}{2};\frac{1-z}{2}+i\,\epsilon\right)\right].
\end{eqnarray}
In the limit when the arguments of the propagator coincide

\be
G_F(1) \propto i\,\cosh(\alpha)\,\frac{\Gamma\left(1-\frac{D}{2} + \omega\right)\,\left|\Gamma(h_+)\right|^2}{\Gamma\left(\frac{D}{2} - 1\right)\, \left|\Gamma\left(\frac12 + i\,\mu\right)\right|^2} + \nonumber \\ \frac{i \,\sinh(\alpha)\, \left|\Gamma\left(h_+\right)\right|^2}{\Gamma\left(\frac{D}{2}\right)\,\Gamma\left(\frac{D}{2}-1\right)}.
\nonumber
\ee
Here $\omega$ is the parameter of the dimensional regularization $D\to D-2\,\omega$ placed only into the divergent expression.

It is straightforward to show that even for general $\alpha$
the integral $\int_{+\infty}^{m^2} dm^2 G_F(1)$ is pure imaginary.
Hence, there is no imaginary contribution to $S_{eff}$. This confirms the conclusions of  \cite{Candelas:1975du,Das:2006wg,Alvarez:2009kq} for the BD propagator
\cite{Bunch:1978yq} and generalizes them for all $\alpha$--vacua.

Consider, instead, such a propagator in dS space, which obeys the composition
principle \cite{Polyakov:2007mm}:

\be
\label{comp}
G_F(z) \propto i\,(z^2 - 1)^{\frac{2-D}{4}}\, Q^{\frac{D-2}{2}}_{-\frac12 + i \mu}(z-2\,i\,\epsilon)
\ee
In this case

\begin{eqnarray}
i\,{\rm Re}\,\int_{+\infty}^{m^2} dm^2 \, G_F(1) \propto {\rm Im}\, \int_{+\infty}^{m^2} dm^2 \, \frac{ \left|\Gamma(h_+)\right|^2}{\Gamma\left(\frac{D}{2}\right)\, \Gamma\left(\frac{D}{2} - 1\right)}\times \nonumber \\ \times \left[F\left(h_+,h_-;\frac{D}{2};1-i\,\epsilon\right) + (-1)^{-h_{-} + 1} \, F\left(h_+,h_-;\frac{D}{2};0+i\,\epsilon\right)\right].\nonumber
\end{eqnarray}
The integral on the RHS of this expression has the usual divergent real contribution,
but its imaginary part is finite:

\be
\frac{{\rm Im} \, L_{eff}}{\sqrt{|g|}} \propto \frac{\pi \, i}{\Gamma\left(\frac{D}{2}\right)\,\Gamma\left(\frac{D}{2}-1\right)}\times \nonumber \\
\int_{+\infty}^{m^2} dm^2 \,\left|\frac12 + i\,\mu\right|^2\dots \left|\frac{D-3}{2} + i\,\mu\right|^2\frac{e^{-\pi\,\mu}}{\cosh(\pi\,\mu)}.\nonumber
\ee
The integration is straightforward and the result in the limit $m^2>>H=1$ is
$ {\rm Im} S_{eff} \propto e^{-2\,\pi\,m}.$
I.e. we clearly see particle production with the Gibbons--Hawking rate \cite{Gibbons:1977mu}
and instability of the vacuum in dS space.
The question is what is the correct in--out Feynman propagator for the problem under consideration?
In the next section we show that it is (\ref{comp}) which follows
from the path integral in dS space and should be the correct in--out Feynman propagator.

\section{Path integral in dS space}

In \cite{Grosche:1987de,Grosche:1987ba} the path integral on the sphere and EAdS space was defined. To define the path integral one represents the Green function as

\be
\label{express}
G(z) = i\, \int_0^\infty dT e^{-\frac{i\,m^2}{2}\, T}\, K\left(z,T\right),
\ee
where the Heat Kernel $K(z=\bar{\eta}_\mu\,\eta^\mu,T)$ solves
$i\,\frac{\partial K\left(z,T\right)}{\partial T} = - \frac12 \, \nabla^2 \, K\left(z,T\right).$
Here $\nabla^2$ is the Laplacian on the $D$--dimensional space in question
and $T$ plays the role of the extra parameter. Initial conditions for
$K(z,T=0)$ define the inhomogeneous part of the Klein--Gordon equation for $G(z)$.

According to \cite{Grosche:1987de,Grosche:1987ba} the path integral for the Feynman propagator
in dS space is:

\be
\label{gz}
G(\bar{\eta}_\mu\, \eta^\mu) \propto \int_0^\infty dT \,
\int_{\eta(0)=\eta}^{\eta(T)=\bar{\eta}} D\eta(t) \,\delta\left[\eta_\mu^2(t) - 1\right]\times \nonumber \\ \times \exp\left[\frac{i}{2}\, \int_0^T dt\left(\dot{\eta}_\nu^2 - m^2 +  \frac{D(D-2)}{4}\right)\right],
\ee
where overdot in $\dot{\eta}$
means differential over the parameter $t$ along the trajectory and the functional
integration goes over all trajectories on dS space: $\eta^2_\mu = - \eta_0^2 + \vec{\eta}^2$.

Naively the expression for the path integral can be obtained both by the analytical continuation
from the sphere ($\eta_0 \to i\,\eta_0$) and from EAdS space ($H\to i\,H$).
The question is which of the analytical continuations gives the proper propagator in dS?
The answer obtained on the sphere \cite{Alvarez:2009kq}, being analytically continued, leads to the BD propagator, while that from EAdS gives the propagator proportional to (\ref{comp}).

Indeed if one were naively analytically continuing from the sphere  to EAdS
($\eta_0 \to i\,\eta_0$ and $H\to i\, H$),
he would have obtained $G(z)= \\ F\left(\mu + \frac{D-1}{2}, -\mu + \frac{D-1}{2};\frac{D}{2}; \frac{1 + z}{2}\right)$, where $\mu = \sqrt{m^2 + (D-1)^2/4}$. However, that is not a correct
propagator in EAdS. The reason is that it is divergent as $z\to \infty$. The
appropriately behaving propagator in EAdS is exactly
$G(z) \propto i\, (z^2 - 1)^{\frac{2-D}{4}}\, Q^{\frac{D-2}{2}}_{-\frac12 + \mu}(z)$ \cite{Burgess:1984ti}.
Such a propagator obeys the composition principle in EAdS.
Under the analytical continuation to dS ($\mu \to i \,\mu$, because $m^2/H^2 \to - m^2/H^2$)
the latter propagator gives (\ref{comp}) rather than (\ref{BD}) with $\alpha=0$.

To calculate the path integral on dS we just have to repeat the calculation of
\cite{Grosche:1987de} with the appropriate analytical continuations performed at every step.
Basically all formulas which are necessary for us are present in \cite{Grosche:1987de} as
equations (31), (34)--(43) of the section II and (1)-(6)
of the section IV. Note the relation between our and their notations:
$d=D+1$, their $m=1$, $R=i$, $u=i\,\eta$, $E = - m^2/2$, $\cosh (r)=z$.
We concisely repeat the main steps of the calculation for the completeness
of our paper.

Before proceeding further let us stress here that if one were doing
analytical continuation from sphere to dS
at each step of the calculation of the path integral \cite{Grosche:1987ba,Alvarez:2009kq}
he would have stuck with the problem that some of the integrals appearing in the course of the calculation
are divergent. The divergence is exactly for the same reason why the BD propagator does not obey the
composition principle \cite{Polyakov:2007mm}. Indeed, to calculate the path integral in question one basically has to use the relation $\int d\eta' \, K(\eta^\nu\,\eta_\nu',T_1)\,K(\bar{\eta}^\gamma\,\eta_\gamma',T_2) = K(\bar{\eta}^\mu\,\eta_\mu,T_1 + T_2)$. The integral on it's LHS becomes divergent (due to the infinite volume of dS
space and due to the presence of the interference terms $e^{-i m\,L(x,y)}\cdot e^{i m\,L(x,y)}$ under the integral) if one first defines it on the sphere and then analytically continues to dS space. At the same time the analytical continuation from EAdS to
dS is completely well defined.

The Heat Kernel can be represented as

\be
\label{lim}
K(\bar{\eta}_\mu\,\eta^\mu,T) = e^{\frac{i\,T\,D\,(D-2)}{8}}\,\lim_{N\to\infty,\,\delta\to 0} \left(\frac{-1}{2\,\pi\,i\,\delta}\right)^{\frac{N\,D}{2}}\nonumber \\
\int \prod_{j=1}^{N-1} d\eta^{(j)}\,
\exp\left\{\frac{i}{\delta}\sum_{j=1}^N\left[1 - \cosh\Delta x^{(j,j-1)}\right]\right\},
\ee
where $\left[\eta^{(j)} - \eta^{(j-1)}\right]^2 = 2\left[1 - \cosh\Delta x^{(j,j-1)}\right]$, and
$x$ are the coordinates on dS: $\eta^2_\mu(x) = 1$.
Then we have to use the decomposition

\be
e^{- c\,\cosh \Delta x} = \sqrt{\frac{2}{\pi\,c}}\, \left[c\,\sinh \Delta x\right]^{\frac{2-D}{2}}\times
\nonumber \\
\int_0^\infty \left|\frac{\Gamma\left(i\,p + \frac{D-1}{2}\right)}{\Gamma(i\,p)}\right|^2\,
P_{-\frac12 + i\,p}^{\frac{2-D}{2}}\left[\cosh \Delta x\right]\, K_{i\,p}(c) \, dp,\nonumber
\ee
where $K_\nu$ is the modified Bessel function, and the relation

\be
(-\sinh \Delta x^{(1,2)})^{\frac{2-D}{2}}\, \left|\frac{\Gamma\left(i\,p + \frac{D-1}{2}\right)}{\Gamma(i\,p)}\right|^2
\, P_{-\frac12 + i\,p}^{\frac{2-D}{2}}(\cosh \Delta x^{(1,2)}) = \nonumber \\ (2\,\pi)^{\frac{D}{2}}\, \sum_{l,n} H^{(D)*}_{p,l,n}(\eta^{(1)})\, H^{(D)}_{p,l,n}(\eta^{(2)}),\nonumber
\ee
where $H^{(D)}_{p,l,n}$ are zonal Spherical Harmonics corresponding to dS,
whose explicit form is not necessary for us. Their orthogonality relation is

\be
\int_{\rm dS} d\eta \, H_{p',l',n'}^{(D)}(\eta)\,H_{p,l,n}^{(D)*}(\eta) = \delta(p-p')\,\delta_{l,l'}\,\delta_{n,n'}.\nonumber
\ee
Using these equations to take the $d\eta^{(j)}$ integrals in (\ref{lim}) and taking the limit $N\to\infty$ with $\delta\to 0$, we get:

\be
K(z,T) = \frac{(z^2 - 1)^{\frac{2-D}{4}}}{(2\,\pi)^{\frac{4-D}{2}}}\, \int_0^\infty dp
\, \left|\frac{\Gamma\left(i\,p + \frac{D-1}{2}\right)}{\Gamma(i\,p)}\right|^2
\times \nonumber \\ P_{-\frac12 + i\,p}^{\frac{2-D}{2}}(z) \, e^{\frac{i\, T}{2}\, \left[p^2 + \left(\frac{D-1}{2}\right)^2\right]}\nonumber
\ee
Substitution of this expression into the integral (\ref{express}), leads to (\ref{comp}).

\section{Acknowledgments}

I would like to thank A.Roura for very valuable discussions
and AEI, Golm, Germany and Subatech, Nantes, France for the hospitality and support
during the time when this work was done. The work was
partially supported by the Federal Agency of Atomic Energy of
Russian Federation and by the grant for scientific schools
NSh-679.2008.2.

\section{Appendix}

In this paper we always consider $D$ to be even.
The homogeneous and isotropic solution of the Einstein equations
$G_{ab} = -\Lambda\, g_{ab},$
with the positive vacuum energy $\Lambda > 0$ is dS space.
Here $g_{ab}$ is the metric tensor with the signature $(-,+,\dots,+)$ and $G_{ab}$ is its Einstein
tensor. The dS space is a
hyperboloid which can be obtained via the analytical
continuation ($\eta_0 \to i\, \eta_0$) from the Euclidian four--sphere (see e.g. \cite{BornerDurr,Kim:2002uz}):

\begin{eqnarray}
\label{sphereeq}
\eta_\mu^2 \equiv - \eta_0^2 + \vec{\eta}^2 = H^{-2},
\end{eqnarray}
where $\Lambda = \frac{(D-1)\,(D-2)}{2}\, H^2$  and $\vec{\eta} = (\eta_1, \dots, \eta_D)$.
Throughout this paper we set $H=1$.

We use here the metric on dS space
which is induced on the hyperboloid through the following solution of
(\ref{sphereeq}) (see e.g.  \cite{BornerDurr,Kim:2002uz});
$\eta_0 = \sinh(t)$, $\eta_i = \cosh(t) \, \omega_i$, where $\sum_i^D \omega_i^2 = 1$
set the coordinates on the unit $(D-1)$--dimensional sphere --- angles $\theta_1,\dots,\theta_{(D-1)}$.
This choice of the coordinates
leads to the global metric on dS: $ds^2 = - dt^2 + \cosh^2(t)\, d\Omega_{(D-1)}^2$, where $d\Omega_{(D-1)}^2$ is the metric on the unit $(D-1)$--dimensional sphere.

Due to $SO(D,1)$ isometry of (\ref{sphereeq}) the Green function $G(\bar{\eta}, \eta)$ in dS space should depend only on the hyperbolic distance $z = \bar{\eta}_\mu\,\eta^\mu$ between the two points $\bar{\eta}_\mu^2 = 1$ and $\eta_\mu^2=1$ in dS space (see e.g. \cite{BornerDurr,Mottola:1984ar,Allen:1985ux,Bousso:2001mw}).
It can be shown that $\nabla_a z \, \nabla^a z = 1 - z^2$ and $\nabla^2_{ab} z = g_{ab} \, z,$
where $\nabla_a$ is the covariant derivative wrt coordinates on dS space $x_a = (t,\theta_1,\dots,\theta_{(D-1)})$. Then the Klein--Gordon
operator, when acting on the dS invariant functions $G(\bar{\eta}, \eta)=G(z)$, can be represented
as

\be
\label{KG}
(\nabla^2 - m^2)\,G(z) \equiv \nonumber \\ \left[(1-z^2)\,\partial_z^2 - D\, z\, \partial_z - m^2\right]\,G(z) = \nonumber \\ A_1\,\partial_z\,\delta(z-1) + A_2\,\partial_z\,\delta(z+1).
\ee
The standard choice of the two--parameter space of Green functions for the Klein--Gordon equation in question is as follows \cite{Mottola:1984ar,Allen:1985ux,Bousso:2001mw}:

\begin{eqnarray}
\label{green}
G(z) = A_1 \, F\left(h_+,h_-;\frac{D}{2};\frac{1+z}{2}\right) + A_2 \, F\left(h_+,h_-;\frac{D}{2};\frac{1-z}{2}\right),
\end{eqnarray}
where $A_{1,2}$ are some numbers, $F$ is the hypergeometric function and

$$
h_{\pm} = \frac{D-1}{2} \pm i\, \mu \quad {\rm and} \quad \mu = \sqrt{m^2 - \left(\frac{D-1}{2}\right)^2}.
$$
Using dimensional regularization ($D\to D-2\omega$) and

\be
F(a,b;c;x) = \frac{\Gamma(c)\,\Gamma(c-a-b)}{\Gamma(c-a)\,\Gamma(c-b)}\, F(a,b;1+a+b-c;1-x)
+ \nonumber \\ \frac{\Gamma(c)\,\Gamma(a+b-c)}{\Gamma(a)\,\Gamma(b)}\, (1-x)^{c-a-b} \, F(c-a,c-b;c-a-b+1;1-x),\nonumber
\ee
where $\Gamma(a)$ is the Gamma--function, one can show that $F\left(h_+,h_-;\frac{D}{2};\frac{1+z}{2}\right)$ is singular at $z=1$,
while $F\left(h_+,h_-;\frac{D}{2};\frac{1-z}{2}\right)$ is singular at $z=-1$. The BD
propagator \cite{Bunch:1978yq}, following from the analytical continuation form the $D$--dimensional
sphere, corresponds (when $m\neq 0$) to

$$A_1 = \frac{i\, \left|\Gamma(h_+)\right|^2}{\Gamma\left(\frac{D}{2}\right)\, \Gamma\left(\frac{D}{2} - 1\right)} \quad {\rm and} \quad A_2 = 0$$ and reproduces the Hadamard short distance
behavior at $z=1$. The following values

\be
A_1 = i\, \cosh(\alpha) \, \frac{\left|\Gamma(h_+)\right|^2}{\Gamma\left(\frac{D}{2}\right)\, \Gamma\left(\frac{D}{2} - 1\right)}\nonumber \\
{\rm and} \quad A_2= i\sinh(\alpha)\, \frac{\left|\Gamma(h_+)\right|^2}{\Gamma\left(\frac{D}{2}\right)\, \Gamma\left(\frac{D}{2} - 1\right)},
\ee
with the real $\alpha$ \cite{Allen:1985ux},
correspond to the dS invariant $\alpha$--vacua (see e.g. \cite{Chernikov:1968zm,Mottola:1984ar,Allen:1985ux,Bousso:2001mw}).

For the case when $A_2\neq 0$ the Green function apart from the standard singularity at $z=1$ (when $\eta_\mu$ resides on the light cone of $\bar{\eta}_\mu$)
has as well a pole at $z=-1$ (when $\eta_\mu$ resides on the light cone of the antipodal point of $\bar{\eta}_\mu$, which is causally disconnected from $\bar{\eta}_\mu$ itself).
See e.g. \cite{Mottola:1984ar,Allen:1985ux,Bousso:2001mw} on the more detailed discussion. I.e. for different $A_1$ and $A_2$ one has different inhomogeneities on the RHS of the Klein--Gordon equation
(\ref{KG}).

Using

\begin{eqnarray}
F(a,b;c;x) = \nonumber \\ \frac{\Gamma(c)\,\Gamma(b-a)}{\Gamma(c-a)\,\Gamma(b)}\,(-x)^{-a}\, F\left(a,a+1-c;a+1-b;\frac{1}{x}\right)
+ \nonumber \\ \frac{\Gamma(c)\,\Gamma(a-b)}{\Gamma(c-b)\,\Gamma(a)}\, (-x)^{-b} \, F\left(b,b+1-c;b+1-a;\frac{1}{x}\right),\nonumber
\end{eqnarray}
one can see that in the limit $z\to\infty$

\be
\label{largz}
F\left(h_+,h_-;\frac{D}{2};\frac{1\pm z}{2}\right) \propto \nonumber \\ \frac{\Gamma\left(\frac{D-1}{2}\right)\,\Gamma(h_- - h_+)}{\Gamma(h_-)\,\Gamma\left(\frac{D-1}{2} - h_+\right)}\, (\mp z)^{-h_+} + \frac{\Gamma\left(\frac{D-1}{2}\right)\,\Gamma(h_+ - h_-)}{\Gamma(h_+)\,\Gamma\left(\frac{D-1}{2} - h_-\right)}\, (\mp z)^{-h_-}.
\ee
The hyperbolic distance $z$ for the large geodesic distance behaves as $z\propto e^L$,
Then for general $A_{1,2}$ the Green function under consideration,
and particularly BD propagator, behaves as $G(z)\propto c\, e^{i\, \mu \, L} + c^*\, e^{-i\, \mu \, L}$ at large $L$ and, hence, does not obey the composition principle \cite{Polyakov:2007mm}.

However, there is another choice of the two--parameter space of Green functions for the Klein--Gordon equation in question \cite{Polyakov:2007mm}

\begin{eqnarray}
G(z) = B_1 \, (z^2 - 1)^{\frac{2-D}{4}}\, P^{\frac{D-2}{2}}_{-\frac12 + i \mu}(z)
+ B_2 \, (z^2 - 1)^{\frac{2-D}{4}}\, Q^{\frac{D-2}{2}}_{-\frac12 + i \mu}(z) \equiv \nonumber \\
C_1\, F\left(h_+,h_-;\frac{D}{2};\frac{1-z}{2}\right)+
C_2\, z^{-h_+}\,F\left(\frac{D+1}{4} + \frac{i\,\mu}{2},\frac{D-1}{4} + \frac{i\,\mu}{2};1+i\,\mu;\frac{1}{z^2}\right), \nonumber
\end{eqnarray}
where $B_{1,2}$ and $C_{1,2}$ are some constants,
$P^n_\nu(z)$ and $Q^n_\nu(z)$ are the associated Legendre functions.
In the case when $B_1=0$ we obtain the Green function, which behaves
in the limit $z\to\infty$ as

\begin{eqnarray}
i\,(z^2 - 1)^{\frac{2-D}{4}}\, Q^{\frac{D-2}{2}}_{-\frac12 + i \mu}(z) \rightarrow {\rm const}\, z^{-h_+}
\nonumber
\end{eqnarray}
and, hence, obeys the composition principle \cite{Polyakov:2007mm}. More generally the Green functions as follows

\begin{eqnarray}
G_{\pm}(z) = \frac{i\,\left|\Gamma(h_+)\right|^2}{\Gamma\left(\frac{D}{2}\right)\, \Gamma\left(\frac{D}{2} - 1\right)} \times \nonumber \\ \left[F\left(h_+,h_-;\frac{D}{2};\frac{1+z}{2}\right) + (-1)^{-h_{\mp} + 1} \, F\left(h_+,h_-;\frac{D}{2};\frac{1-z}{2}\right)\right],\nonumber
\end{eqnarray}
do have proper behavior ($\propto z^{-h_{\pm}}$) in the limit $z\to\infty$, obey the composition principle and have proper Hadamard behavior at $z=1$. However, they as well have poles at $z=-1$.
In particular $G_+(z)\propto i\,(z^2 - 1)^{\frac{2-D}{4}}\, Q^{\frac{D-2}{2}}_{-\frac12 + i \mu}(z)$. Here  $(-1)^{-h_{\mp}+1} = i\,(-1)^{1+\frac{D}{2}}\,e^{\mp \pi \,\mu}$, because D is always taken to be even in this paper.


\begin{thebibliography}{99}


\bibitem{Candelas:1975du}
  P.~Candelas and D.~J.~Raine,
  Phys.\ Rev.\  D {\bf 12}, 965 (1975).

\bibitem{Dowker:1975xj}
  J.~S.~Dowker and R.~Critchley,
  Phys.\ Rev.\  D {\bf 13}, 224 (1976).

\bibitem{Dowker:1975tf}
  J.~S.~Dowker and R.~Critchley,
  Phys.\ Rev.\  D {\bf 13}, 3224 (1976).

\bibitem{Polyakov:2007mm}
  A.~M.~Polyakov,
  arXiv:0709.2899 [hep-th].

\bibitem{Das:2006wg}
  A.~K.~Das and G.~V.~Dunne,
  Phys.\ Rev.\  D {\bf 74}, 044029 (2006)
  [arXiv:hep-th/0607168].

\bibitem{Alvarez:2009kq}
  E.~Alvarez and R.~Vidal,
  arXiv:0907.2375 [hep-th].

\bibitem{Mottola:1984ar}
  E.~Mottola,
  Phys.\ Rev.\  D {\bf 31}, 754 (1985).

\bibitem{Tsamis:1996qq}
  N.~C.~Tsamis and R.~P.~Woodard,
  Nucl.\ Phys.\  B {\bf 474}, 235 (1996)
  [arXiv:hep-ph/9602315].

\bibitem{Tsamis:1992xa}
  N.~C.~Tsamis and R.~P.~Woodard,
  Commun.\ Math.\ Phys.\  {\bf 162}, 217 (1994).

\bibitem{Tsamis:1992zt}
  N.~C.~Tsamis and R.~P.~Woodard,
  Phys.\ Lett.\  B {\bf 292}, 269 (1992).

\bibitem{Tsamis:1993ub}
  N.~C.~Tsamis and R.~P.~Woodard,
  Class.\ Quant.\ Grav.\  {\bf 11}, 2969 (1994).

\bibitem{Tsamis:1994ca}
  N.~C.~Tsamis and R.~P.~Woodard,
  Annals Phys.\  {\bf 238}, 1 (1995).

\bibitem{Tsamis:1996qk}
  N.~C.~Tsamis and R.~P.~Woodard,
  Phys.\ Rev.\  D {\bf 54}, 2621 (1996)
  [arXiv:hep-ph/9602317].

\bibitem{Tsamis:1996qm}
  N.~C.~Tsamis and R.~P.~Woodard,
  Annals Phys.\  {\bf 253}, 1 (1997)
  [arXiv:hep-ph/9602316].

\bibitem{Dolgov:1994cq}
  A.~D.~Dolgov, M.~B.~Einhorn and V.~I.~Zakharov,
  Phys.\ Rev.\  D {\bf 52}, 717 (1995)
  [arXiv:gr-qc/9403056].

\bibitem{Dolgov:1994ra}
  A.~D.~Dolgov, M.~B.~Einhorn and V.~I.~Zakharov,
  Acta Phys.\ Polon.\  B {\bf 26}, 65 (1995)
  [arXiv:gr-qc/9405026].

\bibitem{Weinberg:2006ac}
  S.~Weinberg,
  Phys.\ Rev.\  D {\bf 74}, 023508 (2006)
  [arXiv:hep-th/0605244].

\bibitem{Weinberg:2005vy}
  S.~Weinberg,
  Phys.\ Rev.\  D {\bf 72}, 043514 (2005)
  [arXiv:hep-th/0506236].

\bibitem{Garriga:2007zk}
  J.~Garriga and T.~Tanaka,
  Phys.\ Rev.\  D {\bf 77}, 024021 (2008)
  [arXiv:0706.0295 [hep-th]].

\bibitem{PerezNadal:2007zz}
  G.~Perez-Nadal, A.~Roura and E.~Verdaguer,
  PoS {\bf QG-PH}, 034 (2007).

\bibitem{PerezNadal:2008ju}
  G.~Perez-Nadal, A.~Roura and E.~Verdaguer,
  Class.\ Quant.\ Grav.\  {\bf 25}, 154013 (2008)
  [arXiv:0806.2634 [gr-qc]].

\bibitem{Akhmedov:2008pu}
  E.~T.~Akhmedov and P.~V.~Buividovich,
  Phys.\ Rev.\  D {\bf 78}, 104005 (2008)
  [arXiv:0808.4106 [hep-th]].

\bibitem{Akhmedov:2009vh}
  E.~T.~Akhmedov and E.~T.~Musaev,
  arXiv:0901.0424 [hep-ph].

\bibitem{Akhmedov:2009be}
  E.~T.~Akhmedov, P.~V.~Buividovich and D.~A.~Singleton,
  arXiv:0905.2742 [gr-qc].

\bibitem{Volovik:2008ww}
  G.~E.~Volovik,
  arXiv:0803.3367 [gr-qc].

\bibitem{Akhmedova:2008dz}
  V.~Akhmedova, T.~Pilling, A.~de Gill and D.~Singleton,
  Phys.\ Lett.\  B {\bf 666}, 269 (2008)
  [arXiv:0804.2289 [hep-th]].

\bibitem{Volovik:2009eb}
  G.~E.~Volovik,
  ``Particle decay in de Sitter spacetime via quantum tunneling,''
   arXiv:0905.4639 [gr-qc].

\bibitem{Gibbons:1977mu}
  G.~W.~Gibbons and S.~W.~Hawking,
  Phys.\ Rev.\  D {\bf 15}, 2738 (1977).

\bibitem{Bunch:1978yq}
  T.~S.~Bunch and P.~C.~W.~Davies,
  Proc.\ Roy.\ Soc.\ Lond.\  A {\bf 360}, 117 (1978).

\bibitem{Allen:1985ux}
  B.~Allen,
  Phys.\ Rev.\  D {\bf 32}, 3136 (1985).

\bibitem{Grosche:1987de}
  C.~Grosche and F.~Steiner,
  Annals Phys.\  {\bf 182}, 120 (1988).

\bibitem{Grosche:1987ba}
  C.~Grosche and F.~Steiner,
  Z.\ Phys.\  C {\bf 36}, 699 (1987).

\bibitem{Burgess:1984ti}
  C.~P.~Burgess and C.~A.~Lutken,
  Phys.\ Lett.\  B {\bf 153}, 137 (1985).

\bibitem{Chernikov:1968zm}
  N.~A.~Chernikov and E.~A.~Tagirov,
  Annales Poincare Phys.\ Theor.\  A {\bf 9}, 109 (1968).

\bibitem{BornerDurr}
G.Borner and H.P.Durr, Il Nuovo Cimento, Vol. LXIV A, No. 3 (1969).

\bibitem{Kim:2002uz}
  Y.~b.~Kim, C.~Y.~Oh and N.~Park,
  arXiv:hep-th/0212326.

\bibitem{Bousso:2001mw}
  R.~Bousso, A.~Maloney and A.~Strominger,
  Phys.\ Rev.\  D {\bf 65}, 104039 (2002)
  [arXiv:hep-th/0112218].

\end{thebibliography}
\end{document}